\documentclass[fleqn,usenatbib]{mnras}

\usepackage{newtxtext,newtxmath}
\usepackage{rotating}
\usepackage{booktabs}
\usepackage{multirow}
\usepackage[T1]{fontenc}
\usepackage{subfigure}
\usepackage{orcidlink}

\usepackage[T1]{fontenc}
\usepackage{graphicx}	% Including figure files
\usepackage{amsmath}	% Advanced maths commands
\usepackage{subcaption}  
\usepackage{color}
\usepackage{xcolor}
\newcommand{\LCDM}{\rm{\Lambda}CDM}
\newcommand{\oCDM}{w\mathrm{CDM}}
\newcommand{\ooCDM}{w_0 w_a\mathrm{CDM}}

\usepackage{soul}

\usepackage[normalem]{ulem}

\DeclareRobustCommand{\VAN}[3]{#2}
\let\VANthebibliography\thebibliography
\def\thebibliography{\DeclareRobustCommand{\VAN}[3]{##3}\VANthebibliography}

\title[cosmological constraints from CSST galaxy-scale lenses]{CSST Strong Lensing Preparation: Cosmological Constraints Forecast from CSST Galaxy-Scale Strong Lensing}

\author[H. Wu et al.]{\parbox{\textwidth}{
Hengyu Wu,$^{1}$
Yun Chen\orcidlink{0000-0001-8919-7409},$^{2,3}$
Tonghua Liu\orcidlink{0000-0002-6717-8810},$^{1}$ 
Xiaoyue Cao \orcidlink{0000-0003-4988-9296},$^{4}$
Tian Li\orcidlink{0009-0005-5008-0381},$^{5}$
Hui Li\orcidlink{0009-0005-9594-8375},$^{4}$
Nan Li\orcidlink{0000-0001-6800-7389},$^{2,3}$
Ran Li\orcidlink{0000-0003-3899-0612},$^{6}$
and 
Tengpeng Xu\orcidlink{0000-0002-9855-2342}$^{7}$
\\}
\\
$^{1}$School of Physics and Optoelectronic, Yangtze University, Jingzhou 434023, China\thanks{E-mail: liutongh@yangtzeu.edu.cn (THL)}\\
$^{2}$National Astronomical Observatories, Chinese Academy of Sciences, Beijing 100101, China\thanks{E-mail: chenyun@bao.ac.cn(YC); nan.li@nao.cas.cn(NL)}\\
$^{3}$College of Astronomy and Space Sciences, University of Chinese Academy of Sciences,
Beijing, 100049, China\\
$^{4}$Institute for Astrophysics, School of Physics, Zhengzhou University, Zhengzhou, 450001, China\\
$^{5}$Institute of Cosmology and Gravitation, University of Portsmouth, Burnaby Rd, Portsmouth PO1 3FX, UK\\
$^{6}$School of Physics and Astronomy, Beijing Normal University,  Beijing 100875, China\\
$^{7}$School of Physics and Astronomy, Sun Yat-Sen University, Zhuhai 519082, China\\
}

\date{Accepted XXX. Received YYY; in original form ZZZ}

\pubyear{\the\year{}}

\begin{document}
\label{firstpage}
\pagerange{\pageref{firstpage}--\pageref{lastpage}}
\maketitle

\begin{abstract}
Strong gravitational lensing by galaxies is a powerful tool for studying cosmology and galaxy structure. The China Space Station Telescope (CSST) will revolutionize this field by discovering up to $\sim$100,000 galaxy-scale strong lenses, a huge increase over current samples. To harness the statistical power of this vast dataset, we forecast its cosmological constraining power using the gravitational-dynamical mass combination method. We create a realistic simulated lens sample and test how uncertainties in redshift and velocity dispersion measurements affect results under ideal, optimistic, and pessimistic scenarios. We find that increasing the sample size from 100 to 10,000 systems dramatically improves precision: in the $\Lambda$CDM model, the uncertainty on the matter density parameter, $\Omega_m$, drops from 0.2 to 0.01; in the $w$CDM model, the uncertainty on the dark energy equation of state, $w$, decreases from 0.3 to 0.04. With 10,000 lenses, our constraints on dark energy are twice as tight as those from the latest DESI BAO measurements. We also compare two parameter estimation techniques—MultiNest sampling and Bayesian Hierarchical Modeling (BHM). While both achieve similar precision, BHM provides more robust estimates of intrinsic lens parameters, whereas MultiNest is about twice as fast. This work establishes an efficient and scalable framework for cosmological analysis with next-generation strong lensing surveys.
\end{abstract}

\begin{keywords}
 gravitational lensing: strong – Galaxies: structure – cosmological parameters –
cosmology: observations.
\end{keywords}

%%%%%%%%%%%%%%%%%%%%%%%%%%%%%%%%%%%%%%%%%%%%%%%%%%

%%%%%%%%%%%%%%%%% BODY OF PAPER %%%%%%%%%%%%%%%%%%

\section{Introduction}
Strong gravitational lensing (SGL), a key prediction of general relativity (GR) \citep{1936Sci....84..506E, 2024SSRv..220...12S}, has developed into an invaluable tool in modern astronomy \citep{2025NatAs...9.1116L,2025RSPTA.38340117S,2025ApJ...986...42W}. Depending on the nature of the foreground lens and background source, SGL systems can be categorized into distinct types. Among these, galaxy-galaxy strong lensing (GGSL) systems—where light from a background galaxy is significantly deflected and distorted by a foreground lens galaxy—produce multiple images or extended arcs \citep{2024SSRv..220...87S}. As an independent cosmological probe, SGL offers unique advantages for constraining cosmological parameters \citep{2005ApJ...622...81M,2008A&A...477..397G,2010MNRAS.405.2579O,2010ApJ...708..750S,2014ApJ...788L..35S,2015JCAP...02..010C,2017MNRAS.465.4914B,2020ApJ...888...32L,2022A&A...668A..51L,2022ChPhL..39k9801L,2022ApJ...927L...1W} and reconstructing the mass distribution of lens objects \citep{1998ApJ...509..561K,2006ApJ...649..599K,2011ApJ...727...96R,2016MNRAS.459.3677L,2019MNRAS.489.2049N,2025RAA....25f5013D}.

Galaxy-scale SGL has emerged as a particularly powerful means to study lens properties and refine cosmological constraints. Three primary statistical approaches are widely used: \textbf{(i) lensing probability statistics} \citep{1984ApJ...284....1T,1992ApJ...384....1K,2003MNRAS.343..639O,2005ApJ...622...81M,2012ApJ...755...31C,2023PDU....4101234L}, which relies on the statistical properties of lens populations, making it less sensitive to systematic errors in individual lens modeling yet highly sensitive to the assumed velocity dispersion function (VDF) of lens galaxies, with accuracy dependent upon a complete and well-understood sample to mitigate observational biases; \textbf{(ii) gravitational-dynamical mass combination} \citep{2001PThPh.105..887F,2006PhRvD..73b3006B,2008A&A...477..397G,2015ApJ...806..185C,2019MNRAS.488.3745C,2020ApJ...898..100W}, based on the GR principle which states that within the Einstein radius, the gravitational mass ($M_{\rm{grl}}^E$) must equal the dynamical mass ($M_{\rm{dyn}}^E$), though its reliability depends critically on dynamical mass estimates assuming equilibrium, spherical symmetry, and constrained orbital anisotropy—violations of which may mimic deviations from GR; \textbf{(iii) time-delay measurements} \citep{1964MNRAS.128..307R,2012ApJ...757...82B,2016A&ARv..24...11T,2020MNRAS.498.1420W,2020A&A...643A.165B,2022A&ARv..30....8T,2025MNRAS.tmp.1634X,2025arXiv250603023T}, which use time delays between images to directly constrain the Hubble constant ($H_0$) independent of the distance ladder, but require high-resolution imaging and long-term monitoring, resulting in relatively small sample sizes and high resource demands. While each method offers distinct advantages, they also introduce specific limitations, underscoring the value of combining multiple approaches in SGL cosmology. Furthermore, the methodological limitations of current samples highlight the need for both larger datasets and more robust analysis frameworks.

The advent of stage IV large-area imaging surveys is poised to address the challenge of sample size \citep{2015ApJ...811...20C}. Facilities like the CSST \citep{2025arXiv250704618C}, Euclid \citep{2022A&A...662A.112E}, and Vera C. Rubin Observatory (Rubin LSST) \citep{2009arXiv0912.0201L} are expected to increase the number of detected galaxy-scale SGL systems by about 2-3 orders of magnitude, presenting unprecedented opportunities for cosmological applications. The CSST \citep{2025SCPMA..6880402G,2024MNRAS.533.1960C,2025RAA....25b5015C,2023MNRAS.519.1132M,2022MNRAS.515.5743L}, equipped with a two-meter aperture and a wide field of view covering approximately 17,500 square degrees, achieves an image resolution comparable to that of the Hubble Space Telescope. Its capability for spectral imaging across seven bands ($u$, $g$, $r$, $i$, $z$, $y$, and $NUV$) provides rich multi-band spectral information, enabling accurate conversion of angular measurements from lensing images into physical quantities. This allows the CSST to perform tasks currently challenging for the Rubin LSST \citep{2009AAS...21346003I,2019ApJ...873..111I}, such as precise decomposition of mass components—both stellar and dark matter—within lens galaxies \citep{2009ApJ...705.1099A}. Furthermore, the CSST is expected to uncover previously undetectable lensing systems, thereby enriching the diversity and completeness of known strong lens populations. By combining multi-band imaging with high spatial resolution, the CSST complements other leading observational facilities and offers essential support for advancements in cosmological research.

However, the anticipated increase in galaxy-scale SGL systems from the current $\sim 10^2$ to $\sim 10^5$ presents a major computational challenge. To develop an efficient pipeline for this large-scale dataset, we systematically compare two powerful parameter estimation techniques—the MultiNest sampling algorithm and Bayesian Hierarchical Modeling (BHM). This comparison, evaluating both computational speed and statistical accuracy, is crucial for selecting the optimal method and forms the foundation for a future software package optimized for large samples.

Using a simulated CSST GGSL catalogue, we then apply this framework to forecast cosmological constraints via the gravitational-dynamical mass combination method. Our analysis specifically investigates how uncertainties in redshift and velocity dispersion measurements propagate into the constraints for the $\Lambda$CDM, $w$CDM, and $w_0w_a$CDM models. This work establishes a robust and scalable framework for harnessing the statistical power of next-generation strong lensing surveys.

The rest of this paper is organized as follows. Section ~\ref{sec:models} introduces the theoretical foundation of our method and the cosmological models. Section ~\ref{sec:data_methods} details the simulated lens sample and the two statistical approaches employed. Our results on cosmological constraints, algorithmic performance, and uncertainty analysis are presented and discussed in Section ~\ref{sec:results}. We conclude with a summary of our findings in Section ~\ref{sec:summary}.

\section{Theoretical Framework and Models}
\label{sec:models}

\subsection{Gravitational-Dynamical Mass Combination Method}

As mentioned previously, three statistical approaches are widely used in cosmological studies of galaxy-scale SGL systems. In this work, we focus on GGSL systems, wherein the lensed sources are galaxies rather than transient objects. Consequently, the time-delay method is not applicable. Furthermore, the method of lensing probability statistics can be susceptible to significant observational biases. A systematic investigation of such biases in the CSST GGSL sample will be conducted in future work. Given these considerations, we adopt the gravitational-dynamical mass combination method in this study.

The gravitational-dynamical mass combination method rests on the fundamental assumption that, within the Einstein radius , the gravitational mass equals the dynamical mass:
\begin{equation} \label{eq:Mgrl_Mdyn}
M_{\rm{grl}}^E = M_{\rm{dyn}}^E,
\end{equation}
where $M_{\rm{grl}}^E$ and $M_{\rm{dyn}}^E$ denote the gravitational and dynamical mass, respectively. This equality is naturally satisfied within the framework of GR.

The gravitational mass $M_{\rm{grl}}^E$, derived from the SGL effect, also depends on cosmological distances via the relation: 
\begin{equation} \label{eq:Mgrl}
M_{\rm{grl}}^E = \frac{c^2}{4G}\frac{D_lD_s}{D_{ls}}\theta_E^2,
\end{equation}
where $c$ is the speed of light, $G$ is the gravitational constant, $\theta_{\mathrm{E}}$  denotes the Einstein angular radius, and $D_{l}$, $D_{s}$, and $D_{ls}$ represent the angular diameter distances from the observer to the lens, the observer to the source, and the lens to the source, respectively.

The dynamical mass $M_{\rm{dyn}}^E$ is derived from the stellar velocity dispersion of the lens galaxy. Its estimation depends not only on cosmological distances but also on the assumed mass distribution model of the lens. As a result, the velocity dispersion is connected to a set of model parameters that include both cosmological parameters and those characterizing the mass profile of the lens galaxy.

In this work, we adopt a general mass model for the lens galaxies with E/S0 morphologies \citep{2006EAS....20..161K}, defined as follows:
\begin{equation}\label{eq:lens_mass_model}
\begin{cases}
\rho(r) =  \rho_0 \left( \frac{r}{r_0} \right)^{-\gamma} \\
\nu(r) = \nu_0 \left( \frac{r}{r_0} \right)^{-\delta} \\
\beta(r) =  1 - \frac{\sigma_\theta^2}{\sigma_r^2}.
\end{cases}
\end{equation}
Here, $\rho(r) = \rho_0 \left(r/r_0 \right)^{-\gamma}$ represents the mass density profile, which includes both baryonic and dark matter components, with $\rho_0$ as the normalization constant, $r_0$ a scale radius, and $\gamma$ the power-law slope governing the radial mass distribution. The stellar luminosity density $\nu(r) = \nu_0 \left(r/r_0\right)^{-\delta}$ describes the spatial distribution of starlight, where $\nu_0$ is the normalization and $\delta$ is the luminosity density slope. The orbital anisotropy parameter $\beta(r) = 1 - \sigma_\theta^2/\sigma_r^2$ quantifies the degree of velocity anisotropy, with $\sigma_\theta$ and $\sigma_r$ denoting the tangential and radial velocity dispersions, respectively. A value of $\beta = 0$ corresponds to isotropic stellar orbits, and $\beta = 1$ indicates fully circular motion. When $\gamma = \delta = 2$ and $\beta = 0$, the model reduces to the Singular Isothermal Sphere (SIS) profile—a key approximation for galaxy lenses. It is generalized by the Singular Isothermal Ellipsoid (SIE), which adds ellipticity for greater realism. In the subsequent statistical analysis, we treat $\delta$ as a nuisance parameter and marginalize over it using a Gaussian prior of $\delta = 2.173 \pm 0.085$. This prior is derived from an independent constraint based on a well-studied sample of lens galaxies with high-resolution \textit{Hubble Space Telescope} (HST) imaging \citep{2019MNRAS.488.3745C}. Throughout this work, unless otherwise specified, the Gaussian prior on $\delta$ is applied over the range $[\overline{\delta} - 2\sigma_{\delta}, \overline{\delta} + 2\sigma_{\delta}]$, where $\overline{\delta} = 2.173$ and $\sigma_{\delta} = 0.085$.

By solving the Jeans equation under the lens mass model given in Eq. (\ref{eq:lens_mass_model}), the dynamical mass can be expressed as:
\begin{equation} \label{eq:Mdyn}
M^E_{\rm{dyn}} = \frac{\sqrt{\pi}}{2G}\sigma^2_{\parallel}(\leq R_A) R_E \times f^{-1}( \gamma, \delta, \beta )
\left( \frac{R_A}{R_E} \right)^{\gamma-2},
\end{equation}
where $\sigma^2_{\parallel}(\leq R_A)$ is the observed line-of-sight velocity dispersion of the lens galaxy, luminosity-weighted and averaged within the effective spectroscopic aperture of radius $R_A$. Here, $R_A = D_l \theta_A$ and $R_E = D_l \theta_E$ denote the physical aperture radius and Einstein radius, respectively. The factor $f(\gamma, \delta, \beta)$ is defined as:
\begin{equation}
\begin{aligned}
f(\gamma ,\delta ,\beta )=&\frac{3-\delta }{(\xi -2\beta )(3-\xi )}\left[ \frac{\Gamma (\frac{\xi -1}{2})}{\Gamma (\frac{\xi }{2})}-\beta \frac{\Gamma (\frac{\xi +1}{2})}{\Gamma (\frac{\xi +2}{2})} \right]\\&\times \frac{\Gamma (\frac{\gamma }{2})\Gamma (\frac{\delta }{2})}{\Gamma (\frac{\gamma -1}{2})\Gamma (\frac{\delta -1}{2})}
\end{aligned}
\end{equation}
where $\Gamma$ denotes the Gamma function and $\xi = \gamma + \delta - 2$.

Combining Eqs. (\ref{eq:Mgrl_Mdyn}), (\ref{eq:Mgrl}), and (\ref{eq:Mdyn}) gives the following expression for the velocity dispersion:
\begin{equation} \label{eq:VD_th_RA}
\sigma^2_{\parallel}(\leq R_A) = \frac{c^2}{2\sqrt{\pi}} \frac{D_s}{D_{ls}} \theta_E \times f( \gamma, \delta, \beta )
\left( \frac{\theta_A}{\theta_E} \right)^{2 - \gamma}.
\end{equation}
A detailed derivation can be found in \cite{2019MNRAS.488.3745C}. In practice, both the observed velocity dispersion values and the corresponding model predictions are corrected to a common physical aperture, such as $\theta_{\rm{eff}}/2$, where $\theta_{\rm{eff}}$ is the half-light radius of the lens galaxy. The luminosity-weighted average of the line-of-sight velocity dispersion $\sigma_{\rm{ap}}$, measured within an aperture $\theta_{\rm{ap}}$, is corrected as follows: 
\begin{equation}\label{eq:VD_obs}
\sigma_{\parallel}^{\rm{obs}} \equiv \sigma_{{\rm{e2}}} = \sigma_{\rm{ap}}[\theta_{\rm{eff}}/(2\theta_{\rm{ap}})]^{\eta},
\end{equation}
where the correction exponent $\eta$ is approximately –0.06. For further discussion on the choice of $\eta$, see \cite{1995MNRAS.276.1341J} and \cite{2019MNRAS.488.3745C}. 
Using Eq. (\ref{eq:VD_th_RA}), the corresponding model-predicted value is corrected as:
\begin{equation} \label{eq:VD_th}
\sigma^{\rm{th}}_{\parallel}(\leq \theta_{\rm{eff}}/2) = \left[\frac{c^2}{2\sqrt{\pi}} \frac{D_s}{D_{ls}} \theta_E \times f( \gamma, \delta, \beta )
\left( \frac{\theta_{\rm{eff}}}{2\theta_E} \right)^{2 - \gamma}\right]^{1/2}.
\end{equation}

\subsection{Baseline Cosmological Models}

The model-predicted velocity dispersion (Eq. \ref{eq:VD_th}) depends on the cosmology solely through the distance ratio $D_s/D_{ls}$. In a spatially flat universe ($\Omega_k = 0$), the angular diameter distances are:
\begin{equation}\label{eq:D_s}
D_s(z_s;\boldsymbol{p},H_0) = \frac{c}{H_0}\frac{1}{1+z_s} \int_0^{z_s} \frac{dz}{E(z; \boldsymbol{p})},
\end{equation}
\begin{equation}\label{eq:D_{ls}}
D_{ls}(z_l, z_s;\boldsymbol{p},H_0) = \frac{c}{H_0}\frac{1}{1+z_s} \int_{z_l}^{z_s} \frac{dz}{E(z; \boldsymbol{p})}.
\end{equation}
Here, $\boldsymbol{p}$ represents the cosmological parameters. The dimensionless Hubble parameter is defined as $E(z) \equiv H(z)/H_0$, where $H(z)$ is the Hubble parameter and $H_0$ denotes its value at the present epoch. Consequently, the distance ratio simplifies to:
\begin{equation}\label{eq:distance_ratio}
\mathcal{D}^{\text{th}}(z_l, z_s; \boldsymbol{p}) = \frac{D_{ls}}{D_s} = \frac{\int_{z_l}^{z_s} dz / E(z; \boldsymbol{p})}{\int_{0}^{z_s} dz / E(z; \boldsymbol{p})},
\end{equation}
as the factors of $H_0$ cancel.

The computation of the distance ratio (Eq. \ref{eq:distance_ratio}) for each SGL system requires specifying a cosmological model to determine $E(z)$. Among various extensions to $\Lambda$CDM model\citep{1988PhRvD..37.3406R,2001LRR.....4....1C,2006IJMPD..15.1753C,2006astro.ph..9591A,2008ARA&A..46..385F,2011PhLB..698..175C,2013FrPhy...8..828L}, we consider the $w$CDM and $w_0w_a$CDM models as reference cases. The expressions for $E(z)$ in the cosmological models under consideration are as follows:
\begin{itemize}
\item $\Lambda$CDM model: Parameter set $\boldsymbol{p} = {\Omega_m}$, with
\begin{equation}
E^2(z) = \Omega_m (1+z)^3 + (1 - \Omega_m).
\end{equation}

\item $w$CDM model: Parameter set $\boldsymbol{p} = ({\Omega_m, w}$ ), with
\begin{equation}
E^2(z) = \Omega_m (1+z)^3 + (1-\Omega_m)(1+z)^{3(1+w)}.
\end{equation}

\item $w_0w_a$CDM model: Parameter set $\boldsymbol{p} = (\Omega_m, w_0, w_a)$, with
\begin{equation}
\begin{aligned}
E^2(z) = \Omega_m (1 + z)^3 + (1 - \Omega_m)(1 + z)^{3(1+w_0+w_a)}\\\times \exp\left(-\frac{3w_a z}{1+z}\right).
\end{aligned}
\end{equation}
\end{itemize}

\section{Data Description and Statistical Approaches}
\label{sec:data_methods}

\subsection{Simulated CSST Galaxy-Galaxy Strong Lensing Sample}
\label{subsec:simulated_sample}

\subsubsection{Sample Construction and Properties}
\label{subsubsec:sample_construction}

To evaluate the cosmological constraints achievable with the CSST, we employ a simulated GGSL catalogue from \citet{2024MNRAS.533.1960C}. This simulation populates the sky with early-type lens galaxies using empirical relations from the SDSS and models background sources with a semi-analytical approach. The detectable lensing systems were selected under CSST observational conditions, assuming an SIE mass distribution for the lenses.

To enhance the physical realism of the mock catalogue, we introduce intrinsic scatter in the lens mass profiles. Specifically, we assign each lens a power-law density slope randomly drawn from a Gaussian distribution with a mean of 2.0 and a standard deviation of 0.16, consistent with constraints from the Sloan Lens ACS (SLACS) Survey \citep{2010ApJ...724..511A}.

The CSST wide-field (WF) survey, covering 17,500 square degrees, is predicted to yield a parent sample of $\sim$160,000 GGSL systems. However, for this proof-of-concept study focused on establishing a robust and scalable analysis pipeline, we utilize a representative sub-sample of 10,000 lenses. This sample size is sufficient to capture the statistical power of the future survey while remaining computationally tractable for our detailed methodological comparisons.
Additionally, as discussed below, the severe challenges in measuring velocity dispersions mean that just several tens of thousands of lens systems can be used in this work.
The redshift and velocity dispersion distributions of this sub-sample are representative of the full parent population, ensuring the generalizability of our findings.

\subsubsection{Observational Uncertainties: Redshift and Velocity Dispersion}
\label{subsubsec:accuracy_redshift_vd}

The precision of cosmological parameter constraints is critically dependent on the accuracy of three key observables: the lens redshift ($z_l$), the source redshift ($z_s$), and the lens velocity dispersion ($\sigma_v$).

The precision of redshift measurements is governed by CSST's unique capability for simultaneous photometric and slitless spectroscopic surveys \citep{2019ApJ...883..203G, 2025arXiv250704618C}. For the typically bright lens galaxies, we expect highly secure spectroscopic redshifts with a relative uncertainty of $\Delta z/(1+z) \sim 0.2\%$–$0.5\%$ \citep{2021ApJ...909...53Z, 2024ApJ...977...69Z}. In contrast, redshifts for the background sources will rely primarily on photometric techniques, leading to a larger characteristic uncertainty of $\Delta z/(1+z) \sim 2\%$–$5\%$ \citep{2018MNRAS.480.2178C, 2022RAA....22b5019C}.

Robust measurements of $\sigma_v$ require high-resolution spectroscopy ($R \gtrsim 1000$). We assume that a significant subset of CSST lenses will be located within the footprint of ground-based spectroscopic surveys like DESI, which can supply the requisite data. However, several systematic challenges persist, including low signal-to-noise ratios for fainter lenses, degeneracies in stellar population template fitting, and contamination from the source galaxy's light. Based on the overall technical parameters of the DESI survey, it is estimated that roughly 50,000 of the final 160,000 lens galaxies will possess velocity dispersion data. Precedent from the SLACS survey demonstrates achieved uncertainties of approximately $10\%$, while larger compilations report a range of $5\%$–$20\%$ \citep{2008ApJ...682..964B, 2009ApJ...705.1099A, 2010ApJ...724..511A, 2019MNRAS.488.3745C}. Given that CSST will probe a generally fainter lens population, we adopt a conservative yet realistic uncertainty range of $\Delta\sigma_v/\sigma_v \sim 5\%$–$20\%$.

To systematically quantify the impact of these observational errors on our cosmological inferences, we define three scenarios representing different tiers of measurement precision:

\begin{itemize}
\item \textbf{Ideal case:} This scenario assumes negligible redshift errors and high-precision velocity dispersion measurements ($\Delta\sigma_v/\sigma_v = 5\%$), representing the theoretical upper limit on constraining power:
\begin{eqnarray}
    \Delta z_l &=& 0.0, \nonumber \\
    \Delta z_s &=& 0.0, \nonumber \\
    \Delta \sigma_v &=& 0.05\,\sigma_v.
\label{eq:ideal_case}
\end{eqnarray}
   
\item \textbf{Optimistic case:} This scenario reflects realistically achievable, high-quality data, incorporating spectroscopic precision for lenses, photometric precision for sources, and a $10\%$ velocity dispersion error:
\begin{eqnarray}
    \Delta z_l &=& 0.002\,(1+z_l), \nonumber \\
    \Delta z_s &=& 0.02\,(1+z_s), \nonumber \\
    \Delta \sigma_v &=& 0.10\,\sigma_v.
\label{eq:optimistic_case}
\end{eqnarray}

\item \textbf{Pessimistic case:} This scenario adopts a more conservative outlook with larger uncertainties for all observables, including a $20\%$ error on velocity dispersion:
\begin{eqnarray}
    \Delta z_l &=& 0.005\,(1+z_l), \nonumber \\
    \Delta z_s &=& 0.05\,(1+z_s), \nonumber \\
    \Delta \sigma_v &=& 0.20\,\sigma_v.
\label{eq:pessimistic_case}
\end{eqnarray}
\end{itemize}

In our simulations, we propagate these uncertainties by perturbing the fiducial (true) values of $z_l$, $z_s$, and $\sigma_v$ with Gaussian noise, where the standard deviation for each observable is defined by the equations above.

\subsection{Statistical Approaches}
The anticipated expansion of galaxy-scale SGL samples from hundreds to hundreds of thousands of systems presents a critical computational challenge: parameter estimation methods must be not only statistically robust but also highly efficient to handle the immense data volume. Our preliminary tests on a sample of $\sim$100 systems revealed an order-of-magnitude difference in computational time between commonly used algorithms, underscoring that algorithmic choice is paramount for scalability. To systematically identify the optimal approach for large-scale cosmological inference, we conduct a comprehensive comparison of two statistical strategies for extracting cosmological parameters from large GGSL samples: a population-mean approach, implemented via the MultiNest sampling algorithm, and a hierarchical Bayesian approach, implemented via Bayesian Hierarchical Modeling (BHM).

In the population-mean scheme, the structural parameters (e.g., the total mass-density slope) of all lenses are constrained to follow exact population-level relations, effectively neglecting intrinsic object-to-object scatter. This approach offers substantial computational efficiency and can yield reasonable results when the assumed mean trends are correctly specified. However, it prevents the propagation of unmodelled lens-to-lens variability into the cosmological posterior, which can lead to underestimated uncertainties.

In contrast, the hierarchical Bayesian framework generalizes this configuration by explicitly introducing intrinsic scatter terms around the population relations. This probabilistic linkage allows individual lenses to vary from the mean, and by marginalizing over per-lens latent parameters, it propagates the full structural diversity into the final inference. This results in more reliable uncertainty quantification and reduced bias, albeit at the cost of increased computational complexity.

We assess the limitations and efficacy of these two methods by applying them to simulated datasets of $\mathcal{O}(10^{4})$ CSST GGSLs, providing crucial insights for future cosmological analyses with large lens samples.

\subsubsection{Bayesian Hierarchical Modeling Approach}

We implement the BHM approach \citep{2025arXiv250913435L} using the NumPyro probabilistic programming language \citep{2019arXiv191211554P}, leveraging its built-in No-U-Turn Sampler (NUTS) for posterior inference. As an adaptive variant of Hamiltonian Monte Carlo, NUTS demonstrates superior efficiency in exploring the high-dimensional and complex parameter space of our hierarchical model compared to traditional MCMC methods. Furthermore, by utilizing the JAX library as its backend, we enable large-scale parallel computation on GPUs. This setup not only significantly accelerates the inference process but also provides the necessary scalability to handle the enormous volume of data expected from future surveys like CSST.

To ensure sampling stability and convergence efficiency, we adopt tailored initialization strategies based on cosmological model complexity. For models with fewer parameters, such as $\Lambda$CDM and $w$CDM, we apply the infer.init\_to\_median(num\_samples=20) strategy, initializing from the median of pre-sampled values to enhance stability. In contrast, for more complex models like $w_0w_a$CDM—characterized by a higher parameter count and strong degeneracies (e.g., between $w_0$ and $w_a$)—we employ a central-value scatter method for chain initialization. This approach avoids starting points distant from the target distribution, thereby preserving the performance of the NUTS sampler.

During sampling, we run 10 chains in parallel, each with 1000 burn-in steps followed by 2000 productive sampling steps. By setting the target acceptance probability to target\_accept\_prob = 0.98, we reduce the number of divergent transitions and improve the reliability of the posterior estimates. With this configuration, all scenarios considered in this study yield well-converged parameter constraints, unless otherwise noted.

Our hierarchical model follows the framework of \citet{2024MNRAS.527.5311L}, with the prior distributions for the parameters specified as follows:
\begin{align*}
\text{Cosmology:} \quad
&\begin{cases}
\Omega_m \sim \mathcal{U}(0,1) \\
w \sim \mathcal{U}(-2,0) \\
w_0 \sim \mathcal{U}(-1.5,1) \\
w_a \sim \mathcal{U}(-2,3) \\
\end{cases}
\end{align*}
and 
\begin{align*}
\text{Parent Lens:} \quad
&\begin{cases}
\gamma \sim \mathcal{U}(1.5,2.5) \\
\sigma_\gamma \sim \mathcal{TN}(0.16,0.5,0,0.4) \\
\beta \sim \mathcal{U}(-1,0.5) \\
\sigma_\beta \sim \mathcal{TN}(0.13,0.5,0.1,0.5) 
\end{cases}
\end{align*}
Here, $\mathcal{U}$ denotes a uniform distribution, and $\mathcal{TN}$ represents a truncated normal distribution. The hyperparameters $\sigma_\gamma$ and $\sigma_\beta$ characterize the intrinsic scatters of the lens parameters $\gamma$ and $\beta$, respectively.

\subsubsection{MultiNest sampling algorithm}
To address the computational challenges posed by future large-scale SGL samples, we employ the MultiNest algorithm \citep{2009MNRAS.398.1601F} as a key component of our comparative analysis. MultiNest implements the nested sampling technique, which transforms the challenging multidimensional evidence integration problem into a more tractable one-dimensional one. Its core innovation lies in clustering active points within multidimensional ellipsoids and sampling from these subspaces, operating on the premise that high-likelihood regions are typically found near already sampled points.

This approach enables the algorithm to: (i) automatically detect and separate multiple modes in the posterior distribution; (ii) compute local evidence for each mode individually, providing a robust foundation for model selection; and (iii) generate equally weighted sample points comparable to Markov chains while efficiently handling parameter spaces with complex degeneracies.

Unlike traditional MCMC methods that can struggle to identify and transition between well-separated maxima, MultiNest provides an efficient solution for exploring multi-modal posterior distributions in moderately high-dimensional problems.

We utilize MultiNest through its Python interface, PyMultiNest \citep{2014A&A...564A.125B,2024ApJ...964L...4C} , ensuring seamless integration into our analysis framework. For all analyses presented in this work, we configure the algorithm with 3000 live points to ensure thorough exploration of the parameter space. The sampling stopping criterion, defined by the evidence tolerance, is set to 0.05. This configuration, combined with MultiNest's inherent support for parallel computing, allows us to fully leverage modern multi-core processors, delivering the computational efficiency required for large-scale applications. Our preliminary tests on a sample of $\sim$100 SGL systems confirmed that MultiNest achieves approximately an order-of-magnitude speed improvement over traditional Emcee sampling \citep{2013PASP..125..306F} while maintaining equivalent accuracy in parameter constraints—a crucial advantage for analyzing upcoming survey data.

\section{Results and Discussions} 
\label{sec:results}
 
 To quantitatively assess the cosmological constraining power of the forthcoming CSST GGSL sample, we analyze a representative subset of 10,000 simulated lenses. This sample size is chosen to balance computational feasibility with the goal of capturing the statistical power anticipated from the full survey. Although the total predicted sample contains up to ~160,000 systems, we focus on 10,000 lenses in this work due to computational constraints and the practical challenges associated with obtaining highly accurate velocity dispersion measurements for a significantly larger set. Our investigation systematically addresses three key aspects: (i) the scaling of parameter constraint precision with increasing sample size, (ii) a comparative evaluation of the computational efficiency and statistical robustness of the MultiNest sampling algorithm versus BHM, and (iii) the impact of observational uncertainties—quantified through our optimistic and pessimistic scenarios—on the final cosmological inferences.

\subsection{Scaling of Cosmological Constraints with Sample Size}

\begin{figure*}
\centering
\begin{minipage}[b]{0.45\textwidth}
\centering
\includegraphics[width=\textwidth]{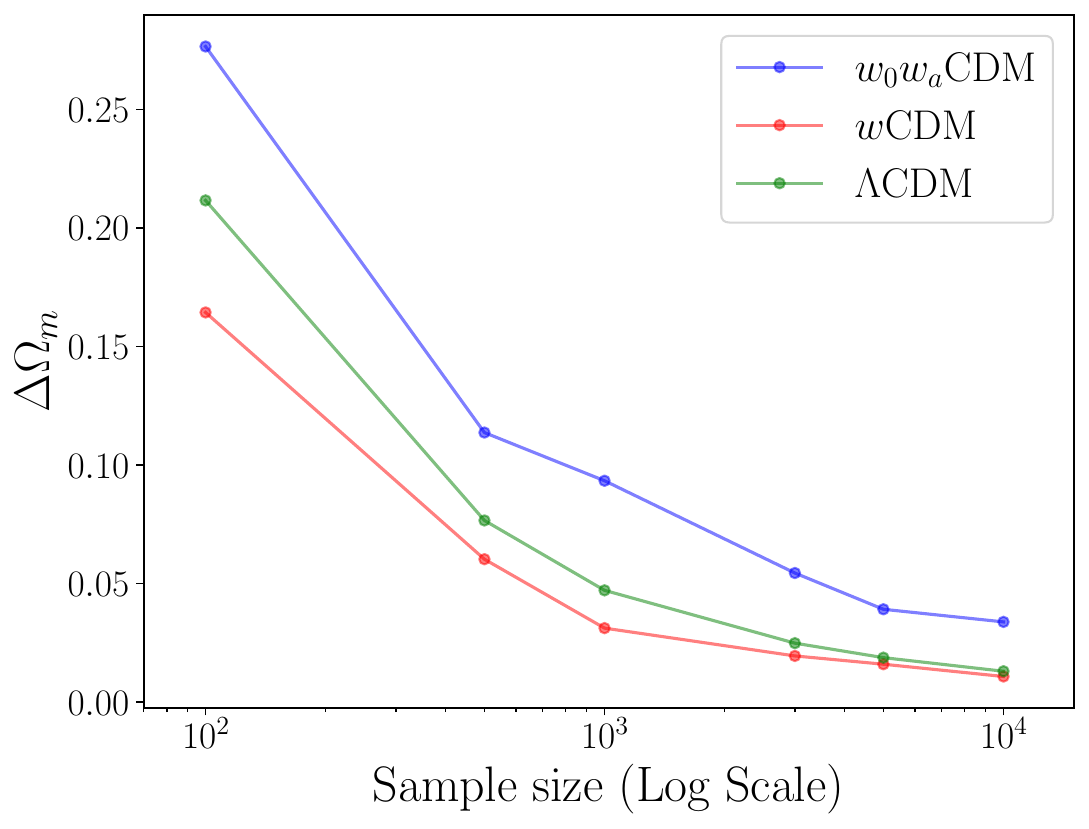}
\end{minipage}
\hfill
\begin{minipage}[b]{0.45\textwidth}
\centering
\includegraphics[width=\textwidth]{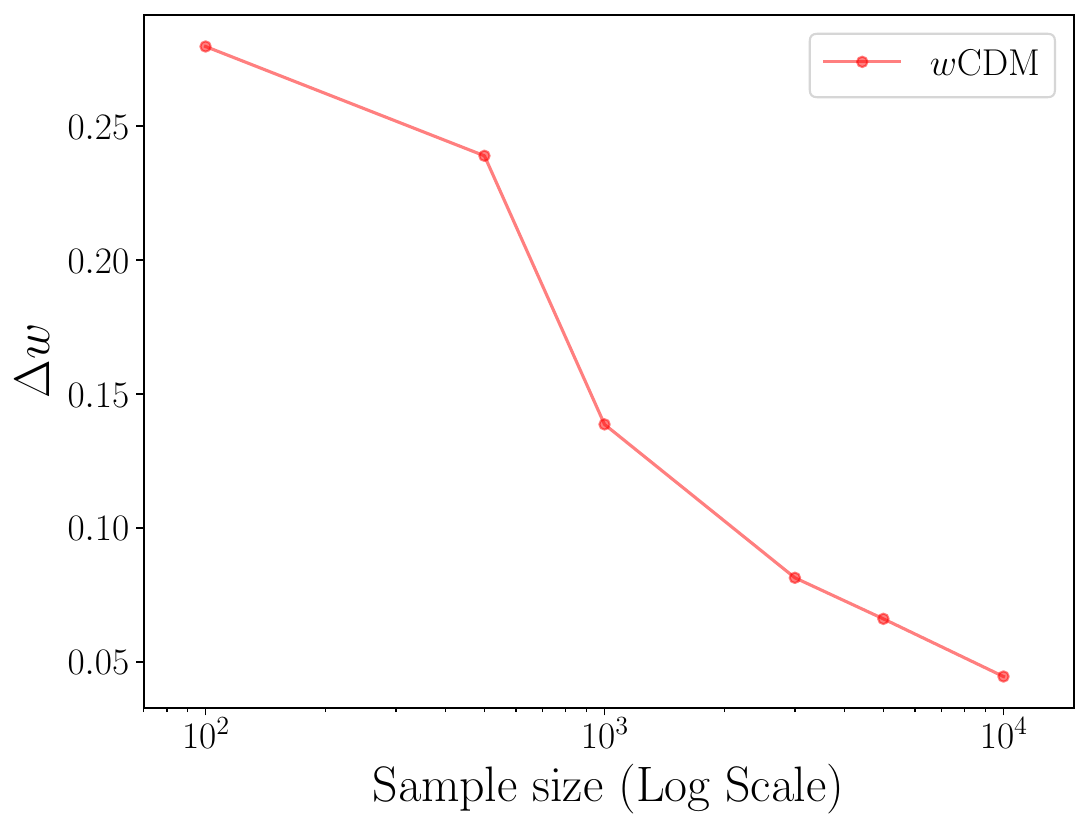}
\end{minipage}
\caption{Dependence of cosmological parameter constraints on the size of the GGSL sample, inferred under the ``\textbf{Ideal case}'' scenario for observational uncertainties in redshifts and velocity dispersion using Bayesian Hierarchical Modeling. The \textbf{left panel} shows the evolution of constraint precision on the matter density parameter $\Omega_m$ for the $\Lambda$CDM, $w$CDM, and $w_0w_a$CDM models. The \textbf{right panel} shows the constraint precision on the dark energy equation of state parameter $w$ in the $w$CDM model.}
\label{fig:Precision_SampleSize}
\end{figure*}

We first quantify the scaling relation between cosmological parameter constraints and the size of the GGSL sample. Figure \ref{fig:Precision_SampleSize} illustrates the dramatic improvement in precision for the matter density parameter $\Omega_m$ and the dark energy equation of state $w$ as the sample size increases from $10^2$ to $10^4$ systems, under the ideal observational scenario (i.e., \textbf{``Ideal case''}) using the BHM approach. The uncertainty on $\Omega_m$ decreases from $\sim 0.20$ to $0.01$ in the $\Lambda$CDM model, while the uncertainty on $w$ drops from $\sim 0.30$ to $0.04$ in the $w$CDM model.

To contextualize the constraining power of the CSST forecast, we compare our results with the latest constraints from DESI BAO measurements \citep{2025PhRvD.112h3515A}. This comparison reveals a key complementarity: for a GGSL sample of $10^4$ systems, the dark energy equation of state is constrained more tightly than with DESI BAO, yielding $w = -1.033^{+0.043}_{-0.046}$ from GGSL versus $w = -0.916 \pm 0.078$ from BAO in the $w$CDM model. In contrast, BAO provides a more precise measurement of the matter density parameter, with $\Omega_m = 0.2975 \pm 0.0086$, compared to $\Omega_m = 0.304^{+0.011}_{-0.015}$ from GGSL in the $\Lambda$CDM model. This establishes galaxy-scale strong lensing as a competitive and complementary probe of dark energy.

\subsection{Computational Trade-offs: MultiNest vs. Bayesian Hierarchical Modeling}

We present a comprehensive comparison of the MultiNest sampling algorithm and Bayesian Hierarchical Modeling (BHM) in terms of computational efficiency and statistical performance. Table \ref{table:comparing_StatisticalApproaches} summarizes the results derived from a sample of $10^4$ GGSL systems under the \textbf{``Ideal case''} scenario.

The analysis yields three principal results. First, MultiNest demonstrates superior computational efficiency, being approximately twice as fast as BHM across all cosmological models considered. Second, despite this difference in speed, both methods produce cosmological parameter constraints of comparable precision. Third, BHM offers greater statistical robustness for characterizing the lens population, as it provides reliable estimates of the intrinsic scatter parameters ($\delta\gamma$ and $\delta\beta$), which are vital for understanding galaxy formation and evolution.

Based on this comprehensive evaluation, we adopted the BHM approach with GPU acceleration for all subsequent analyses. This decision prioritizes the robust handling of intrinsic scatter and population-level parameter estimation, which is crucial for cosmological analyses where understanding systematic uncertainties is paramount.

\begin{table*}
\renewcommand\arraystretch{1.8}
\caption{Comparison of computational efficiency and statistical performance between the MultiNest sampling algorithm and Bayesian Hierarchical Modeling (BHM). Results are derived from a sample of $10^4$ galaxy-galaxy strong lensing systems under the \textbf{``Ideal case''} scenario.} \label{table:comparing_StatisticalApproaches}
\centering
\scalebox{0.9}{
\begin{tabular}{ l| l |l| c c c c c c c}

\hline
\hline
Model & Algorithm & Running Time &$\Omega_m$  & $w$ or $w_0$ & $w_a$ & $\gamma$ & $\delta\gamma$ & $\beta$ & $\delta\beta$
\\
\hline
\multirow{2}{*}{$\LCDM$}
\,\, & MultiNest (CPU) & 0.21 hrs & \,\,$0.302^{+0.015}_{-0.014}$ & --- & --- & $2.001^{+0.003}_{-0.003}$ & --- & $0.181^{+0.012}_{-0.013}$ & ---\\
     & Bayesian hierarchical (GPU) & 0.36 hrs & \,\,$0.304^{+0.011}_{-0.015}$ & --- & --- & $2.004^{+0.003}_{-0.003}$& $0.158^{+0.003}_{-0.002}$ & $0.168^{+0.010}_{-0.010}$ &$0.128^{+0.005}_{-0.006}$\\
\hline
\multirow{2}{*}{$\oCDM$}
& MultiNest (CPU) & 0.38 hrs &\,\,$0.288^{+0.018}_{-0.017}$ & $-1.068^{+0.056}_{-0.056}$ & ---  & $2.001^{+0.003}_{-0.003}$ & --- & $0.193^{+0.016}_{-0.016}$ & --- \\
& Bayesian hierarchical (GPU) & 0.63 hrs &\,\,$0.287^{+0.010}_{-0.011}$ & $-1.033^{+0.043}_{-0.046}$ & ---  &$2.001^{+0.003}_{-0.003}$ & $0.165^{+0.003}_{-0.003}$ & $0.186^{+0.009}_{-0.008}$ & $0.125^{+0.004}_{-0.005}$\\
\hline
\multirow{2}{*}{$\ooCDM$}
& MultiNest (CPU)  & 0.61 hrs &\,\,$0.240^{+0.049}_{-0.059}$ & $-1.185^{+0.150}_{-0.115}$& $0.894^{+0.487}_{-0.846}$  & $2.000^{+0.003}_{-0.003}$ & --- & $0.201^{+0.017}_{-0.018}$ & --- \\
& Bayesian hierarchical (GPU) & 0.85 hrs & \,\,$0.287^{+0.062}_{-0.034}$ & $-0.920^{+0.196}_{-0.233}$& $-0.299^{+0.763}_{-1.181}$  & $1.998^{+0.004}_{-0.004}$ & $0.162^{+0.004}_{-0.004}$ & $0.181^{+0.018}_{-0.017}$ & $0.161^{+0.007}_{-0.007}$ \\
\hline
\hline
\end{tabular}
}
\end{table*}

\subsection{Systematic Uncertainties: The Role of Redshift and Velocity Dispersion Errors}

We systematically evaluate the impact of observational uncertainties on cosmological inferences by comparing the \textbf{``Optimistic case''} and \textbf{``Pessimistic case''} scenarios. Table \ref{table:Optimistic_Pessimistic_cases} summarizes the constraint results from a $10^4$ GGSL sample using the BHM approach.

Our analysis reveals a direct correlation between data quality and cosmological precision. Specifically, the \textbf{``Optimistic case''} not only runs approximately twice as fast as the \textbf{``Pessimistic case''}, indicating enhanced numerical stability with higher-quality data, but also yields parameter constraints that are about twice as precise. Furthermore, the challenge is particularly pronounced for the complex $w_0w_a$CDM model, in which the observational uncertainties under the pessimistic scenario prevent the successful convergence of parameters.

The failure to constrain the $w_0w_a$CDM model under the \textbf{``Pessimistic case''} scenario highlights a fundamental limitation: large redshift errors induce strong degeneracies in the high-dimensional parameter space and can produce non-physical predictions, causing the likelihood function to collapse. This demonstrates that precise redshift measurements are not merely desirable but essential for constraining dynamic dark energy models with strong lensing. Rather than attempting algorithmic corrections post hoc, the most effective strategy is to prioritize the acquisition of high-quality redshift data with lower intrinsic uncertainties.

\begin{table*}
\renewcommand\arraystretch{1.8}
\caption{Comparison of cosmological constraints under \textbf{``Optimistic case''} and \textbf{``Pessimistic case''} scenarios for observational uncertainties, derived from a $10^4$ GGSL sample using Bayesian Hierarchical Modeling, where ``None'' indicates that convergent values could not be obtained.}\label{table:Optimistic_Pessimistic_cases}
\centering
\scalebox{0.9}{
\begin{tabular}{ l| l |l| c c c c c c c}

\hline
\hline
Model & Case & Running Time &$\Omega_m$  & $w$ or $w_0$ & $w_a$ & $\gamma$ & $\delta\gamma$ & $\beta$ & $\delta\beta$
\\
\hline
\multirow{2}{*}{$\LCDM$}
\,\, & Optimistic  & 0.25 hrs & \,\,$0.323^{+0.015}_{-0.020}$ & --- & --- & $2.011^{+0.004}_{-0.003}$ & $0.124^{+0.002}_{-0.003}$ & $0.172^{+0.014}_{-0.011}$ & $0.101^{+0.001}_{-0.001}$\\
     & Pessimistic & 0.59 hrs & \,\,$0.293^{+0.032}_{-0.038}
$ & --- & --- & $1.999^{+0.008}_{-0.006}$ & $ 0.160^{+0.006}_{-0.006}$ & $0.202^{+0.027}_{-0.023}$ &$0.155^{+0.020}_{-0.015}$\\
\hline
\multirow{2}{*}{$\oCDM$}
& Optimistic & 0.29 hrs &\,\,$0.305^{+0.018}_{-0.023}$ & $-1.129^{+0.088}_{-0.083}$ & ---  & $2.012^{+0.004}_{-0.003}$ & $0.124^{+0.002}_{-0.003}$ & $ 0.182^{+0.016}_{-0.012}$ & $0.101^{+0.000}_{-0.001}$ \\
& Pessimistic & 0.69 hrs &\,\,$0.322^{+0.043}_{-0.046}$ & $-0.668^{+0.171}_{-0.183}$ & ---  &$1.995^{+0.008}_{-0.007}$ & $0.158^{+0.006}_{-0.006}$ & $0.168^{+0.033}_{-0.027}$ & $0.163^{+0.020}_{-0.016}$\\
\hline
\multirow{2}{*}{$\ooCDM$}
& Optimistic  & 0.83 hrs &\,\,$0.252^{+0.074}_{-0.041}$ & $-1.194^{+0.105}_{-0.245}$& $0.698^{+1.035}_{-0.407}$  & $2.012^{+0.004}_{-0.004}$ & $ 0.124^{+0.002}_{-0.002}$ & $0.187^{+0.015}_{-0.013}$ & $ 0.101^{+0.001}_{-0.001}$ \\
& Pessimistic & None & \,\,None & None & None  & None & None & None & None \\
\hline
\hline

\end{tabular}
}
\end{table*}

\subsection{Parameter Degeneracies}

\begin{figure*}
\centering
\includegraphics[width=0.80\textwidth]{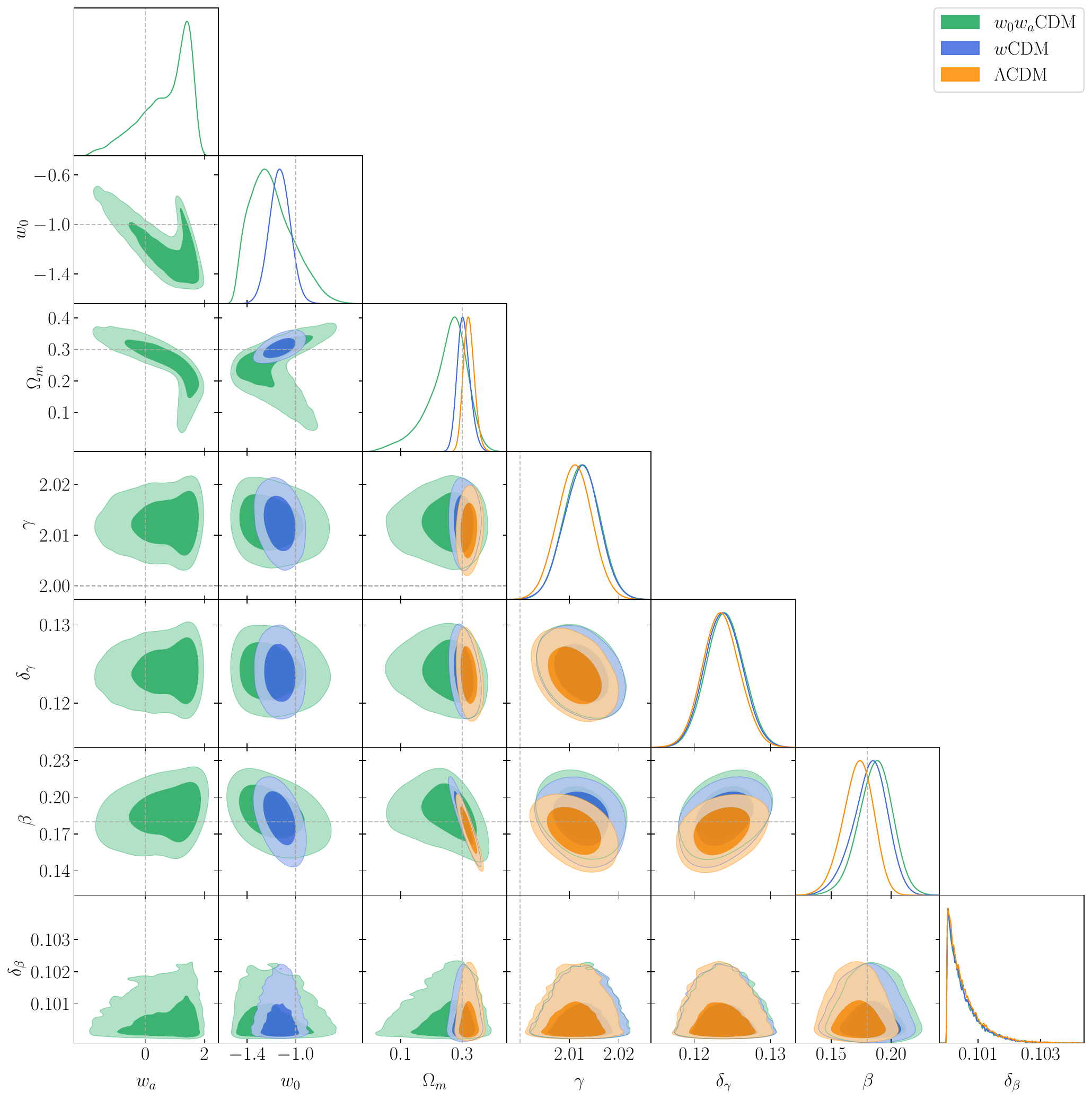}
\caption{One- and two-dimensional posterior distributions of cosmological and lens parameters from 10,000 GGSL systems, inferred under the \textbf{``Optimistic case''} scenario using Bayesian Hierarchical Modeling. Contours enclose the 68\% and 95\% confidence levels. The grey dashed lines indicate the benchmark input values. Cosmological models are color-coded: $\Lambda$CDM (orange), $w$CDM (blue), and $w_0w_a$CDM (green).}
\label{fig:contours}
\end{figure*}

To understand the covariance between parameters and identify potential sources of systematic bias, we examine the degeneracies among cosmological and lens model parameters. Figure \ref{fig:contours} shows the one- and two-dimensional probability distributions for the $\Lambda$CDM, $w$CDM, and $w_0w_a$CDM models, derived from 10,000 GGSL systems under the \textbf{``Optimistic case''} scenario using BHM.

The posterior distributions reveal several notable degeneracies: (i) among cosmological parameters, $w_0$ and $\Omega_m$ are positively correlated, while $w_0$ and $w_a$ show a strong negative degeneracy; (ii) the lens parameters $\gamma$ and $\beta$ are strongly anti-correlated, consistent with the known degeneracy in joint lensing and dynamical analyses; (iii) the anisotropy parameter $\beta$ exhibits significant correlations with cosmological parameters, implying that mismodeling of the stellar orbital structure could bias cosmological results, whereas the mass-density slope $\gamma$ shows no such degeneracies with cosmology.

These degeneracy patterns emphasize the importance of the hierarchical approach, which properly marginalizes over the intrinsic distributions of lens parameters, thereby providing more robust cosmological constraints.

\section{Conclusions}
\label{sec:summary}
The next generation of wide-field surveys, particularly the CSST, is poised to revolutionize strong lensing cosmology by discovering orders of magnitude more galaxy-scale strong lensing systems. In this work, we have performed a comprehensive forecast of the cosmological constraints achievable with the anticipated CSST GGSL sample, utilizing the gravitational-dynamical mass combination method. We have also conducted a critical comparison of parameter estimation techniques to address the computational challenges associated with these large datasets.

Our analysis leads to the following key conclusions:

\begin{itemize}
\item \textbf{Significant Cosmological Constraining Power:} The CSST GGSL sample exhibits tremendous potential to deliver stringent constraints on fundamental cosmological parameters, especially the matter density parameter $\Omega_m$ and the dark energy equation of state $w$. We demonstrate that increasing the sample size from the current $\sim 10^2$ to $\sim 10^4$ systems improves the precision on $\Omega_m$ and $w$ by more than an order of magnitude. With 10,000 lenses, our forecasts yield constraints on $w$ that are about a factor of two tighter than those from the latest DESI BAO measurements, underscoring the unique and complementary value of GGSL for probing dark energy.

\item \textbf{Critical Impact of Data Quality:} The precision of cosmological constraints is highly sensitive to the quality of the input data, specifically the uncertainties in the redshifts of lenses and sources, and the velocity dispersion of the lenses. Our analysis of \textbf{Ideal}, \textbf{Optimistic}, and \textbf{Pessimistic} scenarios clearly indicates that reduced observational uncertainties lead to significantly tighter parameter constraints. A concerted effort to obtain high-S/N spectroscopy for velocity dispersion measurements and precise photometric redshifts for sources will be crucial for maximizing the scientific return from the CSST lens sample.

\item \textbf{Computational Trade-offs: MultiNest vs. Hierarchical Modeling:} For large-scale samples, the choice of parameter estimation algorithm is critical. Our systematic comparison shows that both the MultiNest sampling algorithm and BHM approach are powerful and viable tools. MultiNest offers a substantial advantage in computational speed, being approximately twice as fast as BHM, which is a significant factor for processing tens of thousands of systems. In contrast, BHM provides a more robust statistical framework for inferring the intrinsic distributions and uncertainties of lens population parameters. The selection between them may thus depend on the specific scientific priorities—whether computational efficiency or comprehensive hierarchical inference is paramount.

\item \textbf{A Robust Framework for Future CSST Strong Lensing Cosmology:} This study establishes a validated and efficient framework for future cosmological analyses with large GGSL samples from CSST. The methodologies developed and compared here are directly applicable to the upcoming era of big data in lensing. Our results strongly motivate the development of a dedicated, optimized software pipeline that leverages the strengths of these advanced statistical techniques to enable accurate and timely cosmological inference.
\end{itemize}

Looking forward, several avenues can further enhance the impact of CSST strong lensing studies. A natural extension involves combining multiple statistical approaches, such as lensing probability statistics, within a unified cosmological analysis framework. Furthermore, a more detailed investigation into systematic uncertainties\citep{2009MNRAS.398..635M,2023A&A...678A...4S}—including the selection function of the CSST lens sample and potential deviations from the assumptions underlying dynamical mass estimates (e.g., spherical symmetry) —will be essential for achieving percent-level precision in cosmology. The arrival of the CSST GGSL catalog will undoubtedly open a new chapter in exploring the fundamental properties of our Universe, and this work provides the necessary foundation to fully exploit its potential.

\section*{Acknowledgements}
Y.C. thanks Prof. Alessandro Sonnenfeld for helpful discussions on the parameter degeneracies. This work was supported by the National Key Research and Development Program of China (Nos.\ 2022YFA1602903 and 2023YFB3002501),  the National Natural Science Foundation of China (Nos.\ 12588202, 12473002, and 12203009), and the China Manned Space Program with grant no.\ CMS-CSST-2025-A03. 

%%%%%%%%%%%%%%%%%%%%%%%%%%%%%%%%%%%%%%%%%%%%%%%%%%
\section*{Data Availability}
The simulated CSST galaxy-galaxy strong lensing population data are available at \url{https://github.com/caoxiaoyue/sim csst lens} and Cao et al. (2024) in \url{https://doi.org/10.1093/mnras/stae1865}. Model posterior chains are available from the corresponding author on request.


\bibliographystyle{mnras}
\bibliography{mnras.bib} % if your bibtex file is called example.bib

\label{lastpage}
\end{document}